\documentclass[twocolumn,showpacs,preprintnumbers,nofootinbib,amsmath,amssymb]{revtex4}

\usepackage{euscript,amssymb,amsmath}

\usepackage{graphicx}
\usepackage{epsfig}

\newcommand{\be}[1]{\begin{equation}\label{#1}}
\newcommand{\ee}{\end{equation}}
\newcommand{\ba}[1]{\begin{eqnarray}\label{#1}}
\newcommand{\ea}{\end{eqnarray}}
\newcommand{\rf}[1]{(\ref{#1})}
\newcommand{\nn}{\nonumber}

\begin{document}

\title{Weak-field limit of $f(R)$-gravity in three and more spatial dimensions}

\author{Maxim Eingorn}\email{maxim.eingorn@gmail.com}  \author{Alexander Zhuk}\email{ai_zhuk2@rambler.ru}

\affiliation{Astronomical Observatory and Department of Theoretical Physics, Odessa National University, Street Dvoryanskaya 2, Odessa 65082, Ukraine}

\begin{abstract} We investigate a point-like massive source in non-linear $f(R)$ theories in the case of arbitrary number of spatial dimensions $D\geq 3$.
If $D>3$ then extra dimensions undergo toroidal compactification. We consider a weak-field approximation with Minkowski and de Sitter background solutions. In both these
cases point-like massive sources demonstrate good agreement with experimental data only in the case of ordinary three-dimensional $(D=3)$ space. We generalize this
result to the case of a perfect fluid with dust-like equations of state in the external and internal spaces. This perfect fluid is uniformly smeared over all extra
dimensions and enclosed in a three-dimensional sphere. In ordinary three dimensional $(D=3)$ space, our formulas are useful for experimental constraints on parameters of
$f(R)$ models.
\end{abstract}

\pacs{04.25.Nx, 04.50.Cd, 04.80.Cc, 11.25.Mj}

\maketitle

%%%%%%%%%%%%%%%%%%%%%%%%%%%%%%%%%%%%%%%%%%%%%%%%%%%%%%%%%%%%%%%%%

\section{\label{sec:1}Introduction}

\setcounter{equation}{0}

The idea of the multidimensionality of our Universe demanded by the theories of unification of the fundamental interactions \cite{Polchinski} is one of the most
breathtaking ideas of theoretical physics. It is very important to suggest experiments which can reveal the extra dimensions. On the other hand, if we can show that
existence of the extra dimensions is in contrast to observations, then these theories are prohibited. This important problem is extensively discussed in recent
scientific literature (see, e.g., \cite{KWE}-\cite{Poplawski}). In our paper \cite{EZ3}, we considered a point-like massive source in Kaluza-Klein models with an
arbitrary number of toroidal internal spaces. The gravitational part of the action was taken in the linear (with respect to the scalar curvature $R$) form. It is well
known that in general relativity in the weak-field limit, this is a good physical approximation
%for astrophysical objects
to describe the famous gravitational experiments (perihelion shift, light deflection and time delay of radar echoes). We expected that in the case of extra dimensions we
can also satisfy the gravitational experiments if the sizes of the internal spaces  will be small enough. To our surprise, it is not the case. We found that the
point-like mass approach contradicts the observational data if the total number of spatial dimensions $D$ is more than three and this result does not depend on sizes of
the internal spaces. Analysis performed in Refs. \cite{EZ4,EZ5} shows that the reason lies in equations of state. In the weak-field limit, point-like massive sources
have dust-like equations of state in all spatial dimensions. However, as it follows from our investigations, for considered Kaluza-Klein models, the concordance with
observations requires non-zero equations of state in the internal spaces. It was shown in Ref. \cite{EZ5} that the only compact astrophysical objects, that satisfy the
observational data with the same accuracy as general relativity, are the latent solitons with a dust-like equation of state in our three dimensions and definite
(non-zero) equations of state in the internal spaces. Moreover, the condition of stability of the internal spaces singles out black strings/branes from the latent
solitons and leads uniquely to $p_i=-\varepsilon/2$ equations of state in the internal spaces, and to the number of the external dimensions $d_0=3$. The main problem
with the black strings/branes is to find a physically reasonable mechanism which can explain how the ordinary particles forming the astrophysical objects (e.g., Sun) can
acquire rather specific equations of state ($p_i=-\varepsilon/2$) in the internal spaces. At the moment, it looks very problematic.

%In the present paper, to circumvent this problem in Kaluza-Klein models, we consider non-linear $f(R)$ theories of gravity.
It is possible that a point-like massive source (with dust-like equation of state in external and internal spaces) does not contradict the observational data in some
other generalized gravitational theories different from the one described in \cite{EZ3}-\cite{EZ5}. In this case we can circumvent the problem mentioned above. To
clarify this question, in the present paper we consider non-linear $f(R)$ theories of gravity for an arbitrary number of spatial dimensions $D\geq 3$. Starting from the
pioneering paper \cite{Star1}, the nonlinear
%(with respect to the scalar curvature $R$)
theories of gravity $f(R)$ have attracted the great deal of interest because these models can provide a natural mechanism of the early inflation. Nonlinear models may
arise either due to quantum fluctuations of matter fields including gravity \cite{BirrDav}, or as a result of compactification of extra spatial dimensions
\cite{NOcompact}. Compared, e.g., to other higher-order gravity theories, $f(R)$ theories are free of ghosts and Ostrogradski instabilities \cite{Woodard}. Recently, it
was realized that these models can also explain the late-time acceleration of the Universe. This fact resulted in a new wave of papers devoted to this topic (see, e.g.,
reviews \cite{review1}-\cite{review4}).

Therefore, in this article we investigate a point-like massive source in non-linear $f(R)$ theories in the case of an arbitrary number of spatial dimensions $D\geq 3$.
In a weak-field approximation these sources result in perturbations of a background metrics. We consider two types of the background metrics: the Minkowski metrics and
the de~Sitter metrics. Concerning the form of the function $f(R)$, we demand only that it should be an analytical function that can be expanded in a Taylor series in the
neighborhood of the background solutions. The main result is that in both these cases point-like massive sources demonstrate good agreement with experimental data only
in the case of ordinary three-dimensional $(D=3)$ space. We generalize this result to the case of a perfect fluid with dust-like equations of state in the external/our
and internal spaces. This perfect fluid is uniformly smeared over all extra dimensions and enclosed in a three-dimensional sphere.
%Despite the negative result for $D> 3$, our formulas are suitable i
In ordinary three dimensional $(D=3)$ space, our formulas are useful for experimental constraints on parameters of $f(R)$ models.

The paper is structured as follows. In Section 2 we consider a weak-field limit of $f(R)$ theories with a point-like mass in the case of the Minkowski background and
demonstrate good agreement with experimental data only for three-dimensional space. The similar result takes place in the case of the de Sitter background (see Section
3). In Appendix, we generalize our investigation to the case of perfect fluid enclosed in a three-dimensional sphere. The main results are summarized in the concluding
Section 4.

%%%%%%%%%%%%%%%%%%%%%%%%%%%%%%%%%%%%%%%%%%%%%%%%%%%%%%%%%%%%%%%%%%%%%%%%%%%%%%%%%%%%%%%%%%%%%%%%%%%%%%%%%%%%%%%%%%%
%%%%%%%%%%%%%%%%%%%%%%%%%%%%%%%%%%%%%%%%%%%%%%%%%%%%%%%%%%%%%%%%%%%%%%%%%%%%%%%%%%%%%%%%%%%%%%%%%%%%%%%%%%%%%%%%%%%

\section{\label{sec:2}Weak-field limit of $f(R)$-gravity. Flat spacetime background}

It is well known (see, e.g., \cite{GMZ1,GMZ2}) that in the case of $f(R)$ gravitational theories with an arbitrary number of spacetime dimensions $\mathcal{D}=1+D\geq
4$, the Einstein equations read
%%%%%%
\ba{2.1} &{}& f'(R)R_{ik}-\frac{1}{2}f(R)g_{ik}-[f'(R)]_{;i;k}\nonumber\\
&{}& +g_{ik}[f'(R)]_{;m;n}g^{mn}=\frac{2S_D\tilde G_{\mathcal{D}}}{c^4}T_{ik}\, , \ea
%%%%%%
where $S_D=2\pi^{D/2}/\Gamma (D/2)$ is the total solid angle (the surface area of the $(D-1)$-dimensional sphere of a unit radius), $\tilde G_{\mathcal{D}}$ is the
gravitational constant in $\mathcal{D}$-dimensional spacetime, the prime denotes differentiation with respect to the scalar curvature $R=R_{ik}g^{ik}$: $\; f'(R)\equiv
df/dR$, and the semicolon $;$ denotes the covariant derivative with respect to the metric coefficients $g_{ik}$. The trace of Eq. \rf{2.1} is
%%%%%
\be{2.2} f'(R)R-\frac{1+D}{2}f(R)+D[f'(R)]_{;m;n}g^{mn}=\frac{2S_D\tilde G_{\mathcal{D}}}{c^4}T\, . \ee
%%%%%
Hereafter, the Latin indices $i,k =0,\ldots ,D$, the Greek indices $\alpha,\beta = 1,\ldots ,D$ and we use the sign convention for the metrics, the Riemann and Ricci
tensors in accordance  with the book \cite{Landau}.

In this section, we assume that in the case of absence of a matter source the spacetime is the Minkowski spacetime: $g_{00}=\eta_{00}=1$, $g_{0\alpha}=\eta_{0\alpha}=0$,
$g_{\alpha\beta}=\eta_{\alpha\beta}=-\delta_{\alpha\beta}$. In presence of matter, the metrics is not a Minkowskian one and we will investigate it in the weak-field
limit. This means that the gravitational field is weak and the velocities of test bodies are small compared to the speed of light $c$. In this case the metrics is only
slightly perturbed from its flat spacetime value:
%%%%%%
\be{2.3} g_{ik}\approx\eta_{ik}+h_{ik}\, , \ee
%%%%%%
where $h_{ik}$ are corrections of the order $1/c^2$. Then, up to the order $1/c^2$, the covariant components of Ricci tensor read \cite{EZ3}:
%%%%%%%
\be{2.4} R_{ik}\approx\frac{1}{2}\left(\frac{\partial^2h_i^l}{\partial x^k\partial x^l}+\frac{\partial^2h_k^l}{\partial x^i\partial x^l}-\frac{\partial^2h_l^l}{\partial
x^i\partial x^k}-\eta^{jl}\frac{\partial^2h_{ik}}{\partial x^j\partial x^l}\right)\, . \ee
%%%%%%%
We are going to investigate the weak-field approximation where the gravitational field is generated by a point-like mass at rest. To hold on the right-hand side of Eqs.
\rf{2.1} and \rf{2.2} the terms up to the order $1/c^2$, components of the energy-momentum tensor are approximated as
%%%%%%
\be{2.5} T_{00}\approx\rho c^2,\ \ \ T_{0\alpha}\approx0,\ \ \ T_{\alpha\beta}\approx0 \ \ \Rightarrow \ \ T=T_{ik}g^{ik}\approx\rho c^2\, , \ee
%%%%%
where the rest mass density is
%%%%%
\be{2.6} \rho\equiv m\delta({\bf r}) \ee
%%%%%
and ${\bf r}=(x^1,x^2,\ldots ,x^D)$ is a $D$-dimensional radius vector.

Now, let us suppose that $f(R)$ is an analytical function and can be expanded in a Taylor series near the flat spacetime background value $R=0$ :
%%%%%%
\be{2.7} f(R)=R+aR^2+o\left(R^2\right),\quad |R|\gg|a|R^2\, , \ee
%%%%%%
where $o\left(R^n\right)$ means that this function contains the terms $R^m$ with powers $m>n$. In Eq. \rf{2.7}, $a\equiv (1/2)f''(0)$ and we assumed that $f(0)=0$. The latter means that the cosmological constant $\Lambda$ is absent in the model in accordance with our
assumption that the background spacetime is flat. We also normalize our model in such a way that $f'(0)=1$ which provides the transition to the usual form of the
Einstein equations in the limit $R \to 0 \; \Rightarrow \; f(R) \to R$. From \rf{2.7} we get
%%%%%
\be{2.8} f'(R)=1+2aR+o(R)\, . \ee
%%%%%
Substituting \rf{2.7} and \rf{2.8} into \rf{2.1}, we obtain (up to the order $1/c^2$)
%%%%%
\be{2.9} R_{ik}-2aR_{;i;k}\approx\frac{2S_D\tilde G_{\mathcal{D}}}{c^4}T_{ik}+\frac{1}{2}R\eta_{ik}-2a\eta_{ik}R_{;m;n}\eta^{mn}\, . \ee
%%%%%
Taking into account that up to the order $1/c^2$
%%%%%%
\be{2.10} R_{;i;k}\approx\frac{\partial^2R}{\partial x^i\partial x^k}\, , \ee
%%%%%%
we can write Eq. \rf{2.9} in the following form:
%%%%
\ba{2.11} &{}&\frac{1}{2}\left(\frac{\partial^2h_i^l}{\partial x^k\partial x^l}+\frac{\partial^2h_k^l}{\partial x^i\partial x^l}-\frac{\partial^2h_l^l}{\partial
x^i\partial
x^k}-\eta^{jl}\frac{\partial^2h_{ik}}{\partial x^j\partial x^l}\right)\nonumber \\
&{}&-2a\frac{\partial^2R}{\partial x^i\partial x^k}\approx\frac{2S_D\tilde
G_{\mathcal{D}}}{c^4}T_{ik}+\frac{1}{2}R\eta_{ik}\nonumber\\
&{}&-2a\eta_{ik}\frac{\partial^2R}{\partial x^m\partial x^n}\eta^{mn}\, , \ea
%%%%%
where we used Eq. \rf{2.4}. With the help of the gauge condition
%%%%%
\be{2.12} \frac{\partial}{\partial x^k}\left(h_i^k-\frac{1}{2}\left(h_l^l+4aR\right)\delta_i^k\right)\approx0\, \ee
%%%%%
the formula \rf{2.11} can be written in the form
%%%%%
\be{2.13} \frac{1}{2}\triangle h_{ik}\approx\frac{2S_D\tilde G_{\mathcal{D}}}{c^4}T_{ik}+\frac{1}{2}R\eta_{ik}+2a\eta_{ik}\triangle R\, , \ee
%%%%%
where $\triangle=\delta^{\alpha\beta}\cfrac{\partial^2}{\partial x^{\alpha}\partial x^{\beta}}$ is the $D$-dimensional Laplace operator and up to the order $1/c^2$ holds
%%%%%%
\be{2.14} \eta^{jl}\frac{\partial^2h_{ik}}{\partial x^j\partial x^l}\approx-\triangle h_{ik},\ \ \ \frac{\partial^2R}{\partial x^m\partial x^n}\eta^{mn}\approx-\triangle
R\, . \ee
%%%%%%
We now turn to Eq. \rf{2.2}. Up to the order $1/c^2$, Eq. \rf{2.2} reads
%%%%%
\be{2.15} -\frac{D-1}{2}R-2aD\triangle R\approx\frac{2S_D\tilde G_{\mathcal{D}}}{c^4}T\, . \ee
%%%%%
To get it, we used Eqs. \rf{2.7}, \rf{2.8}, \rf{2.10} and \rf{2.14}. Combining Eqs. \rf{2.13} and \rf{2.15}, we obtain
%%%%%%
\be{2.16} \frac{1}{2}\triangle h_{ik}\approx \frac{2S_D\tilde G_{\mathcal{D}}}{c^4}\left(T_{ik}-\frac{1}{D-1}T\eta_{ik}\right)-\frac{2a}{D-1}\eta_{ik}\triangle R\, . \ee
%%%%%%
It is worth noting that the weak-field equations \rf{2.13}, \rf{2.15} and \rf{2.16} are valid for any energy-momentum tensor $T_{ik} \sim O(c^2)$.  Eq. \rf{2.16} shows
that for a dust-like perfect fluid \rf{2.5} (with an arbitrary form of the mass density $\rho=\rho({\bf r})$) we have
%%%%%
\ba{2.17}
h_{00}&\approx&\frac{2\phi}{c^2}-\frac{4a}{D-1}R\, ,\quad h_{0\alpha}=0\, ,\nn\\
h_{\alpha\beta}&\approx&\left(\frac{1}{D-2}\, \frac{2\phi}{c^2}+\frac{4a}{D-1}R\right)\delta_{\alpha\beta}\, , \ea
%%%%%
where the function{\footnote{\label{phi}From the expression for $h_{00}$ in \rf{2.17}, it is clear that the function $\phi$ defines only a part of the non-relativistic
gravitational potential. Another part follows from the $R$-term. It originates from the additional scalar degree of freedom of non-linear $f(R)$ theories.}} $\phi$
satisfies the $D$-dimensional Poisson equation
%%%%%
\be{2.18} \triangle\phi=S_DG_{\mathcal{D}}\rho \ee
%%%%%
if the $D$-dimensional gravitational constants $\tilde G_{\mathcal{D}}$ and $G_{\mathcal{D}}$ are related as follows:
%%%%%%%
\be{2.19} G_{\mathcal{D}}=2\tilde G_{\mathcal{D}} \frac{D-2}{D-1}\, . \ee
%%%%%%%
The solutions of the $D$-dimensional Poisson equation \rf{2.18} in the case of a point-like mass source were obtained in \cite{EZ1,EZ2}. To solve \rf{2.18} we should
specify the topology of space and the boundary conditions. As for the boundary conditions, we require that the non-relativistic gravitational potential $\varphi ({\bf
r})$ should have the newtonian behavior $\sim 1/r_3$ , where $r_3$ is the length of a radius vector in three-dimensional space,  at far distances from the gravitating
body. To be more precise, we require the following boundary condition:
%%%%%
\be{2.20} \lim\limits_{r_3\rightarrow+\infty}(r_3h_{00})=-r_g\, , \ee
%%%%%%
where $r_g=2G_Nm/c^2$ is the three-dimensional Schwarzschild radius of the gravitating body with the mass $m$ and $G_N$ is the Newtonian gravitational constant. In the
case of non-compact extra dimensions, i.e. when the $(D=3+d)$-dimensional space has topology $M_D=\mathbb{R}^{3+d}$, the function $\phi$ behaves as $\sim 1/r^{1+d}$,
where $r$ is the length of a radius vector in $D$-dimensional space \cite{EZ1,EZ2}. Obviously, it contradicts the boundary condition \rf{2.20}. Therefore, we suppose
that the $(D=3+d)$-dimensional space has the factorizable geometry of a product manifold $M_D=\mathbb{R}^3\times T^{d}$. Here $\mathbb{R}^3$ describes the
three-dimensional flat external (our) space and $T^{d}$ is a torus which corresponds to a $d$-dimensional internal space with volume $V_d=\prod_{i=1}^d a_{i}$, where
$a_{i}$ are the periods of tori. For this topology and with the boundary condition that at infinitely large distances from the gravitating body the potential must go to
the Newtonian expression, we can find the exact solution of the Poisson equation \rf{2.18} \cite{EZ1,EZ2}. The boundary condition requires that the multidimensional
$G_{\mathcal{D}}$ and Newtonian $G_N$ gravitational constants are connected by the following condition:
%%%%%%
\be{2.21} S_D G_{\mathcal{D}}/V_d =4\pi G_N\, . \ee
%%%%%%
Assuming that we consider the gravitational field of the gravitating mass $m$ at distances much greater than periods of tori, we can restrict ourselves to the zero
Kaluza-Klein mode. For example, this approximation is very well satisfied for the planets of the solar system because the inverse-square law experiments show that the
extra dimensions in Kaluza-Klein models should not exceed submillimeter scales \cite{new}. Then, the solution of Eq. \rf{2.18} reads
%%%%
\be{2.22} \phi({\bf r})\approx -\frac{G_N m}{r_3} = -\frac{r_gc^2}{2r_3}\, . \ee
%%%%
In the case when the gravitating mass is uniformly smeared over the extra dimensions, the approximate formula \rf{2.22} becomes the exact equality \cite{EZ1,EZ2}.

Let us turn now to Eq. \rf{2.15} for the scalar curvature $R$. For a point-like mass, this equation takes the form
%%%%%
\be{2.23} \triangle R-\mu^2 R \approx-\frac{S_D\tilde G_{\mathcal{D}} m}{aDc^2}\delta({\bf r})\equiv -\beta \delta({\bf r})\, , \ee
%%%%%
where
%%%%%
\be{2.24} \mu\equiv \left(-\frac{D-1}{4aD}\right)^{1/2}>0\, . \ee
%%%%%
Obviously, $\mu$ plays the role of a mass for the additional scalar degree of freedom in $f(R)$-gravities. To have a physically reasonable (stable) solutions, we should
demand $\mu^2>0$. It happens for $a=(1/2)f''(0)<0$. Some papers require the opposite sign for $f''(R_0)$. However, it can be easily realized that the choice of this sign
depends on the sign convention for the metrics and the definition of the Riemann tensor. The authors in Refs. \cite{CS1,CS2,BG} have chosen the sign convention similar to
ours and also found that $f''(R_0=0)$ should be negative. In $(D=3)$-dimensional space we get
%%%%%
\be{2.25} \mu^2 = \frac{1}{3|f''(0)|}\, , \ee
%%%%%
which coincides with the well known expression for the mass of the scalar degree of freedom in the case of the background solution $R_0=0$ and the normalization
$f'(0)=1$ (see, e.g., \cite{FL-T,Faraoni}).

We consider first the case when space has topology $M_D=\mathbb{R}^{3+d}$, i.e. the extra dimensions are non-compact. Then, the decreasing solution (i.e.
$\lim\limits_{r\rightarrow+\infty}R=0$) of Eq. \rf{2.23} in the region $r>0$ is
%%%%%%
\be{2.26} R\approx CK_{\frac{D}{2}-1}(\mu r)r^{1-\frac{D}{2}}\ \ \stackrel{r\to+\infty}{\longrightarrow}\ \ C\left(\frac{\pi}{2\mu}\right)^{1/2}r^{\frac{1-D}{2}}e^{-\mu
r}\, , \ee
%%%%%%
where $K_{(D/2)-1}$ is the modified Bessel function of the second kind and $C$ is the constant of integration. In particular case $D=3$, we get $R\approx CK_{1/2}(\mu
r_3)r_3^{-1/2}\rightarrow C\left(\pi/(2\mu)\right)^{1/2}e^{-\mu r_3}/r_3$ that gives the Yukawa contribution to the Newtonian gravitational potential. It can be easily
seen that in the case of non-compact extra dimensions $(D>3)$, neither function $\phi$ nor function $R$ provide the boundary condition \rf {2.20}. Therefore, it is
necessary to consider space with topology $M_D=\mathbb{R}^3\times T^{d}$ and compact extra dimensions. For this topology, it is natural to impose periodic boundary
conditions: $R({\bf r}_3, \xi_1,\xi_2,\ldots ,\xi_i,\ldots ,\xi_d)=R({\bf r}_3, \xi_1,\xi_2,\ldots ,\xi_i+a_i,\ldots ,\xi_d),\, i=1,\ldots ,d$, where $a_i$ denotes a
period in the direction of the extra dimension $\xi_i$. Then it is convenient to rewrite Eq. \rf{2.23} as follows:
%%%%%
\ba{2.27} &{}&\triangle_3 R+\sum\limits_{j=1}^{d}\frac{\partial^2R}{\partial\xi_j^2}-\mu^2R\approx-\beta\delta({\bf r}_3)\prod\limits_{j=1}^{d}\delta(\xi_j),\nonumber\\
&{}&\delta(\xi_j)=\frac{1}{a_j} \sum\limits_{k=-\infty}^{+\infty}\cos\left(\frac{2\pi k}{a_j}\xi_j\right)\, . \ea
%%%%%
We seek a solution in the form
%%%%%%
\ba{2.28} &{}&R\approx\prod\limits_{j=1}^{d}\frac{1}{a_j}\sum\limits_{k_1=-\infty}^{+\infty}...\sum\limits_{k_{d}=-\infty}^{+\infty}f_{k_1...k_{d}}({\bf r}_3)\nonumber\\
&{}& \times\cos\left(\frac{2\pi k_1}{a_1}\xi_1\right)...\cos\left(\frac{2\pi k_{d}}{a_{d}}\xi_{d}\right)\, , \ea
%%%%%%
where the function $f_{k_1...k_{d}}({\bf r}_3)$ satisfies the equation
%%%%%
\be{2.29} \triangle_3f_{k_1...k_{d}}-\left[\sum\limits_{j=1}^{d}\left(\frac{2\pi k_j}{a_j}\right)^2+\mu^2\right]f_{k_1...k_{d}}=-\beta\delta({\bf r}_3)\, . \ee
%%%%%
The only solution of this equation with the boundary condition $\lim\limits_{r_3\rightarrow+\infty}f_{k_1...k_{d}}({\bf r}_3)=0$ is
%%%%%
\ba{2.30} &{}&f_{k_1...k_{d}}({\bf r}_3)=\frac{\beta}{4\pi r_3}\nonumber\\
&{}& \times\exp\left[-\left(\sum\limits_{j=1}^{d}\left(\frac{2\pi k_j}{a_j}\right)^2+\mu^2\right)^{1/2}r_3 \right]\, . \ea
%%%%
Taking into account that for $r_3\to +\infty$ the zero mode $k_j=0$, $j=1,...,d$ gives the main contribution, we obtain at large distances the following asymptote
%%%%%
\be{2.31} R\rightarrow\prod\limits_{j=1}^{d}\frac{1}{a_j}\frac{\beta}{4\pi r_3}e^{-\mu r_3}=\frac{D-1}{4aD(D-2)}e^{-\mu r_3}\, \frac{r_g}{r_3}\, , \ee
%%%%%
where we took into account the relation \rf{2.21} between the Newtonian $G_N$ and multidimensional $G_{\mathcal{D}}$ gravitational constants.

It is worth noting that in the case of the smeared extra dimensions (i.e. when the gravitating mass is uniformly smeared over all extra dimensions), Eq. \rf{2.27} is
reduced to
%%%%%
\be{2.32} \triangle_3 R-\mu^2R\approx-\beta\delta({\bf r}_3){\prod\limits_{j=1}^{d}\frac{1}{a_j}}\, , \ee
%%%%%
and the only solution of this equation with the boundary condition $\lim\limits_{r_3\rightarrow+\infty}R=0$ is the function \rf{2.31}.

Thus, the substitution of expressions \rf{2.22} and \rf{2.31} in Eq. \rf{2.17} results in the following form of $h_{00}$ and $h_{\alpha\beta}$:
%%%%%
\ba{2.33}
h_{00}&\approx&-\frac{r_g}{r_3}\left(1+\frac{1}{D(D-2)}e^{-\mu r_3}\right)\, ,\\
\label{2.34} h_{\alpha\beta}&\approx&-\frac{r_g}{r_3}\left(\frac{1}{D-2}-\frac{1}{D(D-2)}e^{-\mu r_3}\right)\delta_{\alpha\beta}\, . \ea
%%%%%%
With the help of the Green function similar expressions were obtained in \cite{CS1} in the particular case $D=3$. As it follows from \rf{2.33}, the non-relativistic
gravitational potential is
%%%%%
\ba{2.35} &{}&\varphi(r_3)= \frac{c^2}{2}h_{00}=-\frac{G_Nm}{r_3}\left(1+\frac{1}{D(D-2)}e^{-r_3/\lambda}\right),\nonumber\\
&{}&\lambda\equiv\left.\frac{1}{\mu}\right|_{D=3}=\left|3f''(0)\right|^{1/2}>0\, . \ea
%%%%%
Thus, the additional scalar degree of freedom results in the Yukawa correction to the non-relativistic gravitational potential. Neither the inverse square law (ISL)
experiments nor the solar system gravitational tests reveal so far such corrections. They give only upper limits on the parameters of the Yukawa interaction. For
example, in the case $D=3$ in the formula \rf{2.35}, the Yukawa parameter $\alpha$ (which is the prefactor in front of the exponent) is equal to $\alpha
=1/(D(D-2))=1/3\approx 10^{-1/2}$ that results in the upper limit \cite{new}
%%%%%
\be{2.36} \lambda \leq \lambda_{\mbox{max}} \approx 7\times 10^{-3}{\mbox{cm}}\quad \Rightarrow \quad |f''(0)|\lesssim 1.6\times 10^{-5}{\mbox{cm}}^2\, . \ee
%%%%%
A similar restriction on $f''(0)$ was also obtained, e.g., in \cite{BG,NJ}.
In the terms of the mass of the scalar degree of freedom it gives $\mu \gtrsim 2\times 10^{-3}$ eV.
%We can also use this estimate to get a restriction on the second derivative of function $f(R)$:
These estimates show that $r_3 \gg \lambda$ for the classical gravitational tests (e.g., deflection of light) in solar system, and we can neglect the Yukawa corrections
at these distances. Therefore, for $D>3$, we arrive at a discrepancy between $h_{00}\approx -r_g/r_3$ and $h_{\alpha\alpha} \approx -(1/(D-2))r_g/r_3$.
%the Yukawa corrections are negligibly small at distances $r_3 >> \lambda$
%and they cannot compensate the discrepancy between $h_{00}$ and $h_{\alpha\alpha}$ for $D>3$.
In linear (with respect to $R$) Kaluza-Klein models with a point-like massive source, exactly this discrepancy leads to a contradiction with the classical gravitational
tests \cite{EZ3}. {\it Therefore, for considered in this section non-linear $f(R)$ models}\footnote{\label{f(R)}We require only that the function $f(R)$ can be expanded
in series \rf{2.7}.}{\it, a point-like massive source demonstrates good agreement with experimental data only in the case of ordinary three-dimensional $(D=3)$ space.}
In Appendix we demonstrate that the similar conclusion holds also in the case of a dust-like spherically symmetric compact gravitating source.

%%%%%%%%%%%%%%%%%%%%%%%%%%%%%%%%%%%%%%%%%%%%%%%%%%%%%%%%%%%%%%%%%%%%%%%%%%%%%%%%%%%%%%%%%%%%%%%%%%%%%%%%%%%%%%%%%%
%%%%%%%%%%%%%%%%%%%%%%%%%%%%%%%%%%%%%%%%%%%%%%%%%%%%%%%%%%%%%%%%%%%%%%%%%%%%%%%%%%%%%%%%%%%%%%%%%%%%%%%%%%%%%%%%%%%

\section{\label{sec:3}De Sitter spacetime background}

According to current observational data, dark energy dominates the energy density in the modern Universe (see, e.g., the lecture \cite{Bean}). There is a number of
theoretical approaches that have been adopted so far to explain the origin of dark energy \cite{Tsu}. It is usually assumed that the Universe at late times enters the de
Sitter stage. It is well known that the de Sitter space is the constant curvature space:
%%%%%
\be{3.1} R_{0ik}=-\frac{2}{D-1}\Lambda g_{0ik}\quad \Longrightarrow \quad R_0=-\frac{2(D+1)}{D-1}\Lambda\, , \ee
%%%%
where $\Lambda$ is the cosmological constant, and according to the observations $\Lambda$ is extremely small at present time: $\Lambda\sim 10^{-57}{\mbox{cm}}^{-2}$.
Consequently, in contrast to the previous section, we suppose that the background space is de Sitter space, rather than Minkowskian one, with a background metrics
$g_{0ik}$ and the background curvature $R_{0}$. Dealing with $f(R)$ theories, we also assume that $R_0$ defines a stable de Sitter fixed point of $f(R)$ theory and
$\Lambda$ is connected with $f(R_0)$ (see Eq. \rf{3.7} below). The formula \rf{3.1} shows that $R_0$ is the same order of magnitude as $\Lambda$ and should be much less
than the curvature $R\equiv R_0+R_1$ in the vicinity of astrophysical objects, i.e., $|R_1| \gg |R_0|$. Note that we still consider the weak-field approximation. Then,
instead of the expansion \rf{2.7} we have
%%%%%
\be{3.2} f(R)= f(R_0)+f'(R_0)R_1 +\frac12 f''(R_0)R_1^2 +o(R_1^2)\, . \ee
%%%%%
We also suppose for expansions $f(R)$ and $f'(R)$ the following inequalities:
%%%%%
\ba{3.3}
&{}&f(R_0)+f'(R_0)R_1\gg\frac{1}{n!}f^{(n)}(R_0)R_1^n,\ n=2,3,...\, ,\nn \\
&{}&f'(R_0)+f''(R_0)R_1\gg\frac{1}{n!}f^{(n+1)}(R_0)R_1^n,\ n=2,3,...\, .\nn\\ \ea
%%%%%%
Taking into account Eqs. \rf{3.2} and \rf{3.3}, we obtain from Eq. \rf{2.1}
%%%%%
\ba{3.4}
&{}&f'(R_0)R_{0ik}+f''(R_0)R_1R_{0ik}+f'(R_0)R_{1ik}\nn\\
&{}&-\frac{1}{2}f(R_0)g_{ik}-\frac{1}{2}f'(R_0)R_1g_{ik}-[f'(R_0)]_{;i;k}\nonumber\\
&{}&-[f''(R_0)R_1]_{;i;k}+g_{ik}[f'(R_0)]_{;m;n}g^{mn}\nn\\
&{}&+g_{ik}[f''(R_0)R_1]_{;m;n}g^{mn}\approx k_{\mathcal{D}}T_{ik}\, , \ea
%%%%%
where $R_1$,$R_{1ik}$ are the differences between total perturbed values $R$,$R_{ik}$ and corresponding background values $R_0$,$R_{0ik}$, and we introduced the notation
$k_{\mathcal{D}}\equiv 2S_D\tilde G_{\mathcal{D}}/c^4$. Here, we do not require that $R_1$, $R_{1ik}$ are defined up to the linear terms $g_{1ik}=g_{ik}-g_{0ik}$. If
necessary, the corresponding linearized expressions can be found, e.g., in \cite{Dick}. In what follows, we consider the case when background value $R_0={\mbox{const}}$.
Without loss of generality, we also suppose that $f'(R_0)=1$. Under these conditions, we get from \rf{3.4}
%%%%%%
\ba{3.5}
&{}&R_{0ik}+f''(R_0)R_1R_{0ik}+R_{1ik}-\frac{1}{2}f(R_0)g_{0ik}\nn\\
&{}&-\frac{1}{2}f(R_0)g_{1ik}-\frac{1}{2}R_1g_{0ik}-f''(R_0)[R_1]_{;i;k}\nonumber\\
&{}&+g_{0ik}f''(R_0)[R_1]_{;m;n}g^{0mn}\approx k_{\mathcal{D}}T_{ik}\, . \ea
%%%%%%
Zero-order approximation gives
%%%%%
\be{3.6} R_{0ik}-\frac{1}{2}f(R_0)g_{0ik}=0 \quad \Longrightarrow \quad R_0=\frac{1+D}{2}f(R_0)\, , \ee
%%%%%
which demonstrates that $f(R_0)$ plays the role of the cosmological constant (cf. with \rf{3.1}):
%%%%%
\be{3.7} f(R_0)\equiv -\frac{4}{D-1}\Lambda\, . \ee
%%%%%
Obviously, the parameter $r_0\equiv -R_0f'(R_0)/f(R_0)$ (defined in \cite{r}) is equal to $r_0=-2$ for $D=3$ what should hold for the de Sitter fixed point. Taking into
consideration Eqs. \rf{3.6}, we rewrite Eq. \rf{3.5} as follows
%%%%%%
\ba{3.8}
&{}&\frac{f''(R_0)R_1R_0}{1+D}g_{0ik}-\frac{R_0}{1+D}g_{1ik}+R_{1ik}\nn\\
&{}&-\frac{1}{2}R_1g_{0ik}-f''(R_0)[R_1]_{;i;k}\nn \\
&{}&+g_{0ik}f''(R_0)[R_1]_{;m;n}g^{0mn}\approx k_{\mathcal{D}}T_{ik}\, . \ea
%%%%%%

Evidently,
%if gravitating bodies are absent,
because of extreme smallness of $\Lambda$, the background metrics can be approximated by the Minkowskian one: $g_{0ik}\approx \eta_{ik}$\footnote{\label{flat}For
example, in the case $D=3$ this approximation works very well at distances $r_3\ll\Lambda^{-1/2}\sim 10^{28}$cm (see, e.g., the formula (5.76) in \cite{BirrDav}).}. For
the same reason, in solar system (e.g., in vicinity of Sun) $|R_0|\ll|R_1|$. Therefore, we can drop two first terms, containing $R_0$, in the left hand side of Eq.
\rf{3.8} and replace $g_{0ik}$ by $\eta_{ik}$. This reduces Eq. \rf{3.8} to Eq. \rf{2.9}. The only difference is that $a$ is equal now to $f''(R_0)/2$.

Let us consider now Eq. \rf{2.2}. In this case, in complete analogy with the derivation of Eq. \rf{3.4}, we obtain
%%%%%
\ba{3.9}
&{}&f'(R_0)R_0+f''(R_0)R_0R_1+f'(R_0)R_1-\frac{1+D}{2}f(R_0)\nn\\
&{}&-\frac{1+D}{2}f'(R_0)R_1+D[f'(R_0)]_{;m;n}g^{mn}\nonumber\\
&{}&+D[f''(R_0)R_1]_{;m;n}g^{mn}\approx \frac{2S_D\tilde G_{\mathcal{D}}}{c^4}T\, . \ea
%%%%%
It can be easily seen that in zero-order approximation we again obtain Eq. \rf{3.6}. Taking into account \rf{3.6}, the normalization $f'(R_0)=1$, the constancy of $R_0$
and the decomposition $g_{ik}=g_{0ik}+g_{1ik}$, we get
%%%%%
\ba{3.10} &{}& [R_1]_{;m;n}g^{0mn}+\left[-\frac{D-1}{2Df''(R_0)}+\frac{1}{D}R_0\right]R_1\nn\\
&{}& \approx \frac{2S_D\tilde G_{\mathcal{D}}}{c^4Df''(R_0)}T\, . \ea
%%%%%
Therefore, the mass squared of the scalar degree of freedom is
%%%%%%
\be{3.11} \mu^2 = -\frac{D-1}{2Df''(R_0)}+\frac{1}{D}R_0\, . \ee
%%%%%%
In the case of three-dimensional space $D=3$ and the condition $f'(R_0)=1$, the similar formula for the mass squared can be found, e.g., in \cite{FL-T,Faraoni,Starob}.
In the approximation $|R_0|\ll |R_1|$, we can drop in Eqs. \rf{3.9}-\rf{3.11} the terms proportional to $R_0$. Then, keeping in mind the relation $g_{0ik}\approx
\eta_{ik}$, we can easily see that Eq. \rf{3.10} is reduced to Eq. \rf{2.15} where $f''(0)$ should be replaced by $f''(R_0)$. Therefore, up to this replacement, the
formulas \rf{2.33} and \rf{2.34} and following from them conclusions are also suitable in the case of the de Sitter background\footnote{\label{R0}It is worth noting that
despite the fact that in our approach the equations with the de Sitter background look similar to the equations with the flat background, the replacement of $f''(0)$ by
$f''(R_0)$ allows us to investigate a class of models with $f''$ divergent at $R_0=0$. For example, in models of the form of $f(R)=
R^{1+\varepsilon}/[(1+\varepsilon)R_0^{\varepsilon}]\, ,\ \varepsilon\ll1$ (see, e.g., Ref. \cite{0611867}) we get $f''(R_0)=\varepsilon/R_0$, and $f''(R_0)$ can be made
arbitrary small for any small non-zero values of $R_0$.}. For example, the ISL experiments require:
%%%%%
\be{3.12} |f''(R_0)|\lesssim 1.6\times 10^{-5}{\mbox{cm}}^2\, . \ee
%%%%%
This estimate indicates that $|f''(R_0)|^{-1}\gtrsim 10^{5}{\mbox{cm}}^{-2} \gg |R_0|\sim \Lambda \sim 10^{-57}{\mbox{cm}}^{-2}$ and we can really drop the terms
proportional to $R_0$ in Eqs. \rf{3.9}-\rf{3.11}.

%The replacement $f''(0) \to f''(R_0)$ gives also a possibility to consider $f(R)$ models which are divergent at $R\to 0$. Let us consider e.g. models of the %type $f(R)=R^{1+\varepsilon}$ where $\varepsilon <1$. First, we renormalize this function in such a way that $f'(R_0)=1$. Second, we also require that %$f(R_0)=2R_0/(1+D)$. Thus, the renormalized function which satisfies these conditions is
%%%%%%
%\be{3.13}
%f(R)=\frac{R^{1+\varepsilon}}{(1+\varepsilon)R_0^{\varepsilon}}+\frac{R_0^{1+\varepsilon}}{R_0^{\varepsilon}}\frac{1-D+2\varepsilon}{(1+\varepsilon)(1+D)}\, .
%\ee
%%%%%
%The second derivative of this function at the background value $R_0$ is
%%%%%%
%\be{3.14}
%f''(R_0)=\frac{\varepsilon}{R_0}\, .
%\ee
%%%%%%
%This formula shows that even for very small value of $R_0$ the second derivative at this point can be finite due to small value of $\varepsilon$. For example, %the estimate \rf{3.12} $|f''(R_0)|= |\varepsilon/R_0|\lesssim 1.6\times 10^{-5}{\mbox{cm}}^2$ for $|R_0|\sim 10^{-57}{\mbox{cm}}^{-2}$ results in extremely %small values for $\varepsilon$: $\varepsilon \lesssim 10^{-61}$.

Therefore, in this section we arrive at the same conclusion as in the previous section: for non-linear $f(R)$ models satisfying the conditions \rf{3.2},\rf{3.3} and with
the de Sitter background, a point-like massive source demonstrates good agreement with experimental data only in the case of ordinary three-dimensional $(D=3)$ space.

%%%%%%%%%%%%%%%%%%%%%%%%%%%%%%%%%%%%%%%%%%%%%%%%%%%%%%%%%%%%%%%%%%%%%%%%%%%%%%%%%%%
\section{Conclusion}

It is well known that a point-like mass is a very good approximation for classical gravitational tests in general relativity with three spatial dimensions. In our paper
we checked this approximation for non-linear $f(R)$ theories for an arbitrary number of spatial dimensions $D\ge 3$. We performed our investigations in a weak-field
limit and considered two forms of the background solutions when a gravitating matter is absent. First, it is the Minkowski metrics. Second, keeping in mind dark energy
in the Universe, it is the de Sitter metrics. Concerning the form of the function $f(R)$, we demand only that it should be an analytical function that can be expanded in
a Taylor series in the neighborhood of the background solutions. For both of these cases we obtained the similar expressions for the perturbations of the metric
coefficients $h_{00}$ and $h_{\alpha\beta}$. This makes it possible to calculate, for example, the deflection of light in solar system. The form of the perturbations
$h_{00}$ and $h_{\alpha\beta}$ exactly indicate that for both background metrics we get the same conclusion: point-like massive sources demonstrate good agreement with
experimental data only in the case of ordinary three-dimensional $(D=3)$ space. We generalize this result to the case of a perfect fluid with dust-like equations of
state in the external/our and internal spaces. This perfect fluid is uniformly smeared over all extra dimensions and enclosed in a three-dimensional sphere. It is not
difficult to see that more complicated form of the internal structure does not change the situation. It follows from the fact that the asymptotic ratio
$h_{\alpha\alpha}/h_{00} =1/(D-2)$ (which is in agreement with the observations only for $D=3$) takes place for an arbitrary form of $\rho(\bf{r})$ (see Eq. \rf{2.17}).
We think that the found contradiction with the observations is a generic property of the Kaluza-Klein models. The point is that in the case of dust-like equations of
state in the external and internal spaces, both point-like and extended gravitating objects result in the mentioned above asymptotic ratio between the metric
perturbations: $h_{\alpha\alpha}/h_{00} = 1/(D-2)$. It happens, e.g., in Kaluza-Klein models with toroidal \cite{EZ3,EZ4,EZ5} and spherical \cite{EZC} compactifications
as well as in considered above $f(R)$ models. It is also clear that the exterior to a black hole may not coincide with the field exterior to a point mass. We can see
this on the example of the soliton solutions in Kaluza-Klein models with toroidal compactification. Here, latent solitons, black strings and black branes do not
contradict the observations (see, e.g., \cite{EZ4,EZ5}). However, in all these cases we get very strange equations of state in the internal space. It is of interest to
obtain an analogue of the black string/brane solutions for multidimensional $f(R)$ models to see what the form of equations of state takes place here. We are going to
undertake such a study in the near future.

In conclusion, we would like to note that in ordinary three dimensional $D=3$ space, the formulas obtained in our paper are useful for experimental constraints on
parameters of $f(R)$ models. For example, we found the formula for the gravitational force between two spheres. This formula can be used for calculation $f''(R_0)$ from
direct experimental measurements.

%%%%%%%%%%%%%%%%%%%%%%%%%%%%%%%%%%%%%%%%%%%%%%%%%%%%%%%%%%%%%%%%%%%%%%%%%%%
%\newpage
\section*{ACKNOWLEDGEMENTS}

This work was supported in part by the "Cosmomicrophysics" programme of the Physics and Astronomy Division of the National Academy of Sciences of Ukraine.

%%%%%%%%%%%%%%%%%%%%%%%%%%%%%%%%%%%%%%%%%%%%%%%%%%%%%%%%%%%%%%%%%%%%%%%%%%
%%%%%%%%%%%%%%

\section{Appendix: Dust-like perfect fluid in a three-dimensional sphere}
\renewcommand{\theequation}{A\arabic{equation}}
\setcounter{equation}{0}

In this appendix we consider the case where a point-like mass is replaced by a dust-like perfect fluid with the energy-momentum tensor \rf{2.5}. We suppose that this
perfect fluid is uniformly smeared over all extra dimensions and, at the same time, enclosed in a three-dimensional sphere of the radius $r_s$. Then, multidimensional
$\rho$ and three-dimensional $\rho_3$ energy densities are related as $\rho_3=\rho V_d$. We also assume that $\rho_3=$ const. Therefore, Eq. \rf{2.18} outside and inside
the sphere, respectively, reads
%%%%%
\ba{a.1} &{}& \triangle_3\phi_{out}=\frac{1}{r_3}\frac{d^2}{d r_3^2}(r_3\phi_{out})=0,\nn\\
&{}& \triangle_3\phi_{in}=\frac{1}{r_3}\frac{d^2}{d r_3^2}(r_3\phi_{in})=4\pi G_N\rho_3\, , \ea
%%%%%
where we take into account the relation \rf{2.21} between multidimensional $G_{\mathcal{D}}$ and Newtonian $G_N$ gravitational constants. These equations have the
following solutions:
%%%%%
\ba{a.2} &{}& \phi_{out}=-\frac{G_Nm}{r_3},\nn\\
&{}& \phi_{in}=\frac{2\pi G_N\rho_3r_3^2}{3}-2\pi G_N\rho_3r_s^2\nn\\
&{}& =\frac{G_Nmr_3^2}{2r_s^3}-\frac{3G_Nm}{2r_s}\, . \ea
%%%%%
To get these solutions, we use the boundary conditions $\phi_{out}(r_3\to +\infty)\to 0$, $|\phi_{in}(0)|<+\infty$, and the matching conditions
$\phi_{out}(r_s)=\phi_{in}(r_s),\; \left. d\phi_{out}/dr_3\right|_{r_s}=\left. d\phi_{in}/dr_3\right|_{r_s}$.

Similarly, Eq. \rf{2.15} outside and inside the sphere takes the form
%%%%%
\ba{a.3}
&{}& \triangle_3R_{out}-\mu^2R_{out}=\frac{1}{r_3}\frac{d^2}{d r_3^2}(r_3R_{out})-\mu^2R_{out}=0\, , \nn \\
&{}& \triangle_3R_{in}-\mu^2R_{in}=\frac{1}{r_3}\frac{d^2}{d r_3^2}(r_3R_{in})-\mu^2R_{in}=4\pi G_N\tilde \rho_3\, , \nn\\ \ea
%%%%%
where the mass squared $\mu^2$ is defined by Eq. \rf{2.24} and we introduce the notation
%%%%%
\be{a.4} \tilde \rho_3 \equiv \tilde \beta \rho_3\, ,\quad \tilde\beta \equiv -\frac{1}{c^2}\frac{D-1}{2aD(D-2)}\, . \ee
%%%%%
Using the boundary and matching conditions similar to conditions for the functions $\phi$, we obtain the following solutions:
%%%%%
\ba{a.5}
&{}& R_{out}=\frac{e^{-\mu r_3}}{r_3}\, \frac{4\pi G_N\tilde \rho_3}{\mu^3}\left[\sinh(\mu r_s)-\mu r_s\cosh(\mu r_s)\right]\nn \\
&{}& =-\tilde\beta \frac{G_Nm}{r_3}e^{-\mu r_3}\, \frac{3 }{\mu^3r_s^3}\left[\mu r_s\cosh(\mu r_s)-\sinh(\mu r_s)\right] \nn\\ \ea
%%%%%%
and
%%%%%
\be{a.6} R_{in}=\frac{4\pi G_N\tilde \rho_3}{\mu^2}\left[\frac{1}{\mu r_3}e^{-\mu r_s}(1+\mu r_s)\sinh(\mu r_3)-1\right]\, . \ee
%%%%%
With the help of the asymptotic formula $x\cosh x-\sinh x\approx x^3/3$ for $x\ll 1$, we can easily see that $R_{out}$ is reduced to Eq. \rf{2.31} for $r_s\to 0$. On the
other hand, up to the prefactor $\tilde \beta$, $R_{in}$ and $R_{out}$ are reduced to $\phi_{in}$ and $\phi_{out}$ \rf{a.2}, respectively, for $\mu \to 0$. Now, we can
get the metric coefficients $h_{00}$ and $h_{\alpha\beta}$ defined in Eq. \rf{2.17}. For example, we obtain outside of the sphere
%%%%%
\ba{a.7}
&{}& h_{00}^{out}=-\frac{r_g}{r_3}\left(1+\frac{1}{D(D-2)}e^{-\mu r_3}\right.\nn\\
&{}& \left.\times\frac{3}{\mu^3r_s^3}[\mu r_s\cosh(\mu r_s)-\sinh(\mu r_s)]\right)\, ,\label{a.7}\\
&{}& h_{\alpha\beta}^{out}=-\frac{r_g}{r_3}\left(\frac{1}{D-2}-\frac{1}{D(D-2)}e^{-\mu r_3}\right.\nn\\
&{}& \left.\times\frac{3}{\mu^3r_s^3}[\mu r_s\cosh(\mu r_s)-\sinh(\mu r_s)]\right)\delta_{\alpha\beta}\label{a.8}\, . \ea
%%%%%
Similar to a point-like massive source, we also arrive at the relation $h_{\alpha\beta}^{out}\approx h_{00}^{out}/(D-2)$ for $\mu r_3\gg 1$.

It is worth noting that the non-relativistic potential $\varphi_{out}(r_3)=(c^2/2)h_{00}^{out}$ coincides formally with the formula (3.7) in \cite{EZ2} where we should
keep the first Kaluza-Klein modes $k=\pm 1$ and take into account that $\mu \equiv 2\pi \chi_{\pm 1}$, $R_2\equiv r_s\, , R_1\equiv 0$. We also need to replace in (3.7)
the Yukawa parameter $\alpha =2$ (for simplicity we consider in (3.7) the case of one extra dimension) by $\alpha =1/(D(D-2))$. We can apply this analogy to another
formulas in \cite{EZ2}. For example, the gravitational force between two spheres (with masses $m_1$ and $m_2$ and radii $r_{s1}$ and $r_{s2}$) is described by the
formula similar to the expression (4.7) in \cite{EZ2}:
%%%%%
\be{a.9} \frac{F(r_3)}{F_N(r_3)} =1+\frac{9}{4}\alpha \left(\frac{\lambda}{r_{s1}}\right)^2\left(\frac{\lambda}{r_{s2}}\right)^2 \frac{r_3}{\lambda}\;
e^{-(r_3-r_{s1}-r_{s2})/\lambda} \ee
%%%%%
where $F_N(r_3)= -G_N m_1m_2/r_3^2$. Obviously, it makes no sense to use this formula for $D>3$ because this case, as shown above, contradicts the observational data.
However, we can apply it to the case $D=3$ where $\alpha=1/3$. Therefore, since $\left.\lambda\right|_{D=3} = \left|3f''(R_0)\right|^{1/2}$, then we can use this formula
for direct experimental measurement of $f''(R_0)$.
%%%%%%%%%%%%%%%%%%%%%%%%%%%%%%%%%%%%%%%%%%%%%%%%%%%%%%%%%%%%%%%%%%%%%%%%%%%%%%%%%%%%%%%%%%%%%%%%%%%%%%%%%%%%%%%%%
%%%%%%%%%%%%%%%%%%%%%%%%%%%%%%%%%%%%%%%%%%%%%%%%%%%%%%%%%%%%%%%%%%%%%%%%%%%%%%%%


\section*{References}

\begin{thebibliography}{99}

%%%%%%%%%%%
\bibitem{Polchinski}
J. Polchinski, {\em String Theory, Volume 2: Superstring Theory and Beyond} (Cambridge University Press, Cambridge, 1998).
%%%%%%%%%%%%
\bibitem{KWE}
D. Kalligas, P.S. Wesson and C.W.F. Everitt, Astrophys. J. {\bf 439}, 548 (1995).
%%%%%%%%%
\bibitem{limOvWesson}
P.H. Lim, J.M. Overduin and P.S. Wesson, J. Math. Phys. {\bf 36}, 6907 (1995).
%%%%%%%
\bibitem{LO}
H. Liu and J. Overduin, Astrophys. J. {\bf 538}, 386 (2000); (arXiv:gr-qc/0003034).
%%%%%%%
\bibitem{LOW}
T. Liko, J.M. Overduin and P.S. Wesson, Space Sci. Rev. {\bf 110}, 337 (2004); (arXiv:gr-qc/0311054).
%%%%%%%
\bibitem{indians}
F. Rahaman, S. Ray, M. Kalam and M. Sarker, Int. J. Theor. Phys. {\bf 48}, 3124 (2009); (arXiv:gr-qc/0707.0951).
%%%%%%
\bibitem{XuMa}
P. Xu and Y. Ma, Phys. Lett. B {\bf 656}, 165 (2007); (arXiv:gr-qc/0710.3677).
%%%%%%%
\bibitem{Poplawski}
N.J. Poplawski, {\em Einstein-Cartan gravity excludes extra dimensions} (2010); (arXiv:hep-th/1001.4324).
%%%%%%%
%%%%%%%%
\bibitem{EZ3}
M. Eingorn and A. Zhuk, Class. Quant. Grav. {\bf 27}, 205014 (2010); (arXiv:gr-qc/1003.5690).
%%%%%%
%%%%%%%%%%%
\bibitem{EZ4}
M. Eingorn and A. Zhuk, Phys. Rev. D {\bf 83}, 044005 (2011); (arXiv:gr-qc/1010.5740).
%%%%%%%%%%%%
\bibitem{EZ5}
M. Eingorn, O. de Medeiros, L. Crispino and A. Zhuk, Phys. Rev. D {\bf 84}, 024031 (2011); (arXiv:gr-qc/1101.3910).
%%%%%%%%%
\bibitem{Star1}
A.A. Starobinsky, Phys. Lett. B {\bf 91}, 99 (1980).
%%%%%%%
\bibitem{BirrDav}
N.D. Birrell and P.C.W. Davies, {\em Quantum Fields in Curved Space} (Cambridge University Press, Cambridge, 1982).
%%%%%%
\bibitem{NOcompact}
S. Nojiri and S.D. Odintsov, Phys. Lett. B {\bf 576}, 5 (2003), (arXiv:hep-th/0307071).
%%%%%%
\bibitem{Woodard}
R.P. Woodard, Lect. Notes Phys. {\bf 720}, 403 (2007), (arXiv:astro-ph/0601672).
%%%%%%%
\bibitem{review1}
S. Nojiri and S.D. Odintsov, Int. J. Geom. Meth. Mod. Phys. {\bf 4}, 115 (2007); (arXiv:hep-th/0601213).
%%%%%%%%
\bibitem{review2}
T.P. Sotiriou and V. Faraoni, Rev. Mod. Phys. {\bf 82}, 451 (2010); (arXiv:gr-qc/0805.1726).
%%%%%%%%%
\bibitem{review3}
S. Capozziello, M. De Laurentis and V. Faraoni, Living Rev. Rel. {\bf 13}, 3 (2010); (arXiv:gr-qc/0909.4672).
%%%%%
\bibitem{review4}
A. De Felice and S. Tsujikawa, {\em f(R) theories} (2010); (arXiv:gr-qc/1002.4928).
%%%%%%
%%%%%%
\bibitem{GMZ1}
U. G\"unther, P. Moniz and A. Zhuk, Phys. Rev. D {\bf 66}, 044014 (2002); (arXiv:hep-th/0205148).
%%%%%%
%%%%%%
\bibitem{GMZ2}
U. G\"unther, P. Moniz and A. Zhuk, Phys. Rev. D {\bf 68}, 044010 (2003); (arXiv:hep-th/0303023).
%%%%%%
%%%%%%%%
\bibitem{Landau}
L.D. Landau and E.M. Lifshitz, {\em The Classical Theory of Fields, Fourth Edition: Volume 2 (Course of Theoretical Physics Series)} (Oxford Pergamon Press, Oxford,
2000).
%%%%%%%%%%%%
%%%%%%%%%%%%
\bibitem{EZ1}
M. Eingorn and A. Zhuk, Phys. Rev. D {\bf 80}, 124037 (2009); (arXiv:hep-th/0907.5371).
%%%%%%%
\bibitem{EZ2}
M. Eingorn and A. Zhuk, Class. Quant. Grav. {\bf 27}, 055002 (2010); (arXiv:gr-qc/0910.3507).
%%%%%%
%%%%%%%
\bibitem{new}
D.J. Kapner, T.S. Cook, E.G. Adelberger, J.H. Gundlach, B.R. Heckel, C.D. Hoyle and H.E. Swanson, Phys. Rev. Lett. {\bf 98}, 021101 (2007); (arXiv:hep-ph/0611184).
%%%%%%%
\bibitem{CS1}
S. Capozziello and A. Stabile, Class. Quant. Grav. {\bf 26}, 085019 (2009); (arXiv:gr-qc/0903.3238).
%%%%%%%
\bibitem{CS2}
S. Capozziello, A. Stabile and A. Troisi, Phys. Lett. B {\bf 686}, 79 (2010); (arXiv:gr-qc/1002.1364).
%%%%%%%
\bibitem{BG}
C.P.L. Berry and J.R. Gair,
{\em Linearized f(R) Gravity: Gravitational Radiation \& Solar System Tests} (2011); (arXiv:gr-qc/1104.0819).
%%%%%%%
\bibitem{FL-T}
V. Faraoni and N. Lanahan-Tremblay, Phys. Rev. D {\bf 77}, 108501 (2008); (arXiv:gr-qc/0712.3252).
%%%%%%
\bibitem{Faraoni}
V. Faraoni, Class. Quant. Grav. {\bf 26}, 145014 (2009); (arXiv:gr-qc/0906.1901).
%%%%%
\bibitem{NJ}
J. N\"af and P. Jetzer, Phys. Rev. D {\bf 81}, 104003 (2010); (arXiv:gr-qc/1004.2014)
%%%%%%
\bibitem{Bean}
R. Bean, {\em TASI Lectures on Cosmic Acceleration} (2010); (arXiv:astro-ph/1003.4468).
%%%%%%%%
\bibitem{Tsu}
S. Tsujikawa, {\em Dark energy: investigation and modeling. Invited review chapter on dark energy for a book "Dark Matter and Dark Energy: a Challenge for the 21st
Century"} (2010); (arXiv:astro-ph/1004.1493); M. Li, X.-D. Li, S. Wang and  Y. Wang, {Dark Energy} (2011); (arXiv:astro-ph/1103.5870).
%%%%%%%
\bibitem{Dick}
R. Dick, Gen. Rel. Grav. {\bf 36}, 217 (2004); (arXiv:gr-qc/0307052); Y. S. Myung, T. Moon and E. J. Son, {\em Stability of f(R) black holes} (2011);
(arXiv:gr-qc/1103.0343).
%%%%%%%
\bibitem{r}
L. Amendola, R. Gannouji, D. Polarski and S. Tsujikawa, Phys. Rev. D {\bf 75}, 083504 (2007); (arXiv:gr-qc/0612180).
%%%%%%%
\bibitem{Starob}
V. Muller, H.-J. Schmidt and A.A. Starobinsky, Phys. Lett. B
{\bf 202}, 198 (1988).
%%%%%%%
\bibitem{0611867}
T. Chiba, T. L. Smith and A. L. Erickcek, Phys. Rev. D {\bf 75}, 124014 (2007); (arXiv:astro-ph/0611867).
%%%%%%%%%
\bibitem{EZC}
A. Chopovsky, M. Eingorn and A. Zhuk, {\em Weak-field limit of Kaluza-Klein models with spherical compactification: problematic aspects}, in preparation (2011).




%%%%%%%%%%%%%%%%%%%%%%%%%%%%%%%%%%%%%%%%%%%%%%%%%%%%%%%%%%%%%%%%%%%%%%%%%%%%%%%%%%%%%%%%%%

\end{thebibliography}
\end{document}